\documentclass[prb,twocolumn,english,twocolumns,superscriptaddress]{revtex4-1}

\usepackage{amsmath, amssymb,graphicx,bbm}

\usepackage{color}
\usepackage{physics}
\usepackage{docmute}
\usepackage[export]{adjustbox}
\usepackage[nottoc,numbib]{tocbibind}
\usepackage{graphicx}
\usepackage{float}
\usepackage{soul} 
\usepackage{ulem}
\usepackage{umoline}
\usepackage[usenames,svgnames]{xcolor}

\usepackage{hyperref}
\hypersetup{
     colorlinks=true,           
     linkcolor=Crimson,          
     citecolor=blue,            
     filecolor=blue,            
     urlcolor=cyan,             
}

\newcommand{\av}[1]{\left\langle#1\right\rangle}
\newcommand{\com}[1]{\left[#1\right]}

\newcommand{\size}{\Lambda}
\newcommand{\EqDef}{\stackrel{\mathrm{def}}{=}}

\newcommand{\Eq}[1]{Eq.~(\ref{#1})}
\newcommand{\Fig}[1]{Fig.~\ref{#1}}
\newcommand{\iRef}[1]{Ref.~\onlinecite{#1}}

\newcommand{\App}[1]{Appendix~\ref{#1}}
\renewcommand{\i}{{\mkern1.5mu \operatorfont i \mkern1.5mu}}

\newcommand{\pvec}[1]{\vec{#1}\mkern2mu\vphantom{#1}}
\begin{document}

\title{Learning the dynamics of open quantum systems from their steady states}

\author{Eyal Bairey}
\email[]{baeyal@gmail.com}
\affiliation{Physics Department, Technion, 3200003, Haifa, Israel}

\author{Chu Guo}
\affiliation{Quantum Intelligence Lab (QI-Lab), Supremacy Future Technologies (SFT), Guangzhou 511340, China}

\author{Dario Poletti}
\affiliation{Science and Mathematics Cluster and EPD Pillar, Singapore University of Technology and Design, 8 Somapah Road, 487372 Singapore}

\author{Netanel H. Lindner}
\affiliation{Physics Department, Technion, 3200003, Haifa, Israel}

\author{Itai Arad}
\affiliation{Physics Department, Technion, 3200003, Haifa, Israel}

\begin{abstract}

Recent works have shown that generic local Hamiltonians can be efficiently inferred from local measurements performed on their eigenstates or thermal states. Realistic quantum systems are often affected by dissipation and decoherence due to coupling to an external environment. This raises the question whether the steady states of such open quantum systems contain sufficient information allowing for full and efficient reconstruction of the system's dynamics.
We find that such a reconstruction is possible for generic local Markovian dynamics. We propose a recovery method that uses only local measurements; for systems with finite-range interactions, the method recovers the Lindbladian acting on each spatial domain using only observables within that domain. We numerically study the accuracy of the reconstruction as a function of the number of measurements, type of open-system dynamics and system size. Interestingly, we show that couplings to external environments can in fact facilitate the reconstruction of Hamiltonians composed of commuting terms.
\end{abstract}

\maketitle

\section{Introduction} The development of quantum simulators and computation devices has rapidly progressed over the last few years \cite{Preskill2018}. These developments span a multitude of physical platforms, including ultracold atoms \cite{Monroe2002, Bloch2008, Gross2017, Bernien2017}, trapped ions \cite{Cirac1995, Blatt2012, Monroe2013}, photonic circuits \cite{Kok2007, Aspuru-Guzik2012, Flamini2019, Takeda2019}, Josephson junction arrays \cite{You2005, Houck2012, Devoret2013, Wendin2017, Neill2018} and more, reaching ever larger complexity. The growth in the complexity of these systems calls for efficient methods to characterize and verify their dynamics. The resources required by these methods, whether classical computations or quantum measurements, should scale polynomially with the number of degrees of freedom in the system. 

An isolated quantum system can be characterized by learning its underlying Hamiltonian. This can be achieved by monitoring the dynamics that the Hamiltonian generates \cite{DaSilva2011,Burgarth2009, DiFranco2009, Shabani2011a, Zhang2014,Wang2015,
DeClercq2016, Sone2017, Wang2018,Granade2012, Wiebe2014,
Wiebe2014a, Wiebe2015, Wang2017}, or by measuring local observables in one of its eigenstates  or thermal states \cite{Rudinger2015, Huber2016, Kieferova2017, Qi2017, Chertkov2018, Greiter2018,  Kappen2018, Turkeshi2019, Bairey2019}. However, realistic quantum systems are never fully isolated. This raises the need for methods to characterize the dynamics of {\it open quantum systems} which are coupled to external environments. 

Previous works have recovered the dynamics of open quantum systems by tracking their time evolution \cite{Chuang1997c, Childs2001, Boulant2003a, Mohseni2006, Howard2006a, Bellomo2009, Bellomo2010, DaSilva2011, Akerman2012, Glickman2013, Zhang2015, Ficheux2018}. However, the possibility of recovering open system dynamics from their \emph{steady states} has not been addressed. 
 We focus on open quantum systems evolving under Markovian and local dynamics, for which the evolution can be described by the Lindblad master equation formalism \cite{Gorini1976, Lindblad1976}:
\begin{align} \label{lind_def}
    \dot{\rho}&=\mathcal{L} \left(\rho \right) =  \\
&= -\i \sum_j \com{H_j, \rho} + \frac{1}{2} \sum_j \left( \com{L_j \rho, L_j^\dagger} + \com{L_j, \rho L_j^\dagger} \right), \nonumber
\end{align} 
where each $H_j$, $L_j$ is a local operator. Throughout this paper, a local operator will be defined as acting on at most $k$
spatially \textit{contiguous} degrees of freedom (e.g. spins). While the
Hamiltonian terms $H_j$ are Hermitian, the $L_j$ operators, known as
the `jump operators', are generically not.  A steady
state $\rho_s$ of $\mathcal{L}$ is defined by 
$\dot{\rho}_s = \mathcal{L}(\rho_s) = 0$. Suppose that we prepare many copies of $\rho_s$ and measure expectation
values of local observables in the state $\rho_s$. Can
$\mathcal{L}$ be recovered using the data obtained from these measurements?

Parameter counting suggests this should be possible. The number of parameters describing a local Lindbladian scales polynomially with the system size, similarly to a Hamiltonian. On the other hand, a quantum state is described by exponentially many parameters. Thus, the steady state of a local Lindbladian may potentially contain sufficient information for inferring the dynamics that generated it. 

However, steady states of Lindbladians differ from eigenstates and thermal states of local Hamiltonians. Every Hamiltonian commutes with the density matrix corresponding to each of its eigenstates $\ketbra{\epsilon_i}{\epsilon_i}$. In contrast, generic Lindbladians have only a single steady state \cite{Evans1977}. Dissipation can cause this unique steady state to be highly mixed, possibly reducing its information content. 
As an extreme example, the steady state of any Lindbladian whose jump operators $L_j$ are all Hermitian is the fully mixed state $\rho \propto \mathbbm{1} $, from which there is no hope to recover the Lindbladian. Does this impose a fundamental difficulty to Lindbladian reconstruction? Or do the steady states of local many-body dissipative dynamics generically  bear clear signatures of the preceding dynamics? Can these dynamics be extracted efficiently and accurately?

In this work, we study this question by providing an efficient algorithm for learning the dynamics of local Lindbladians from their steady states. Extending the methods of \iRef{Bairey2019}, our algorithm exploits strong constraints that locality imprints on the steady states of generic local Lindbladians. Using this algorithm, we (i) explore which types of Lindbladians can be accurately reconstructed from their steady states, (ii) study numerically and analytically the system-size scaling of the reconstruction accuracy, and (iii)
show that coupling to a bath can in fact facilitate the reconstruction of certain classes of Hamiltonians, which pose a challenge for methods based on their eigenstates or Gibbs states.

\section{Algorithm} We begin by choosing  a basis of local Hermitian operators for the unitary dynamics $\lbrace{h_i \rbrace}$, and a basis of local operator \emph{pairs} for the dissipative dynamics $\lbrace{ \left( l_r, l_s \right) \rbrace}$. Expanding the dynamics in this operator basis (see \App{app:expansion}), \Eq{lind_def} becomes
\begin{equation}
\dot{\rho} = -\i \sum_i c_i \com{h_i, \rho} + \sum_{r,s} \frac{c_{rs}}{2}\left( \com{l_r \rho, l_s^\dagger} + \com{l_r, \rho l_s^\dagger}\right),
\label{lind_expansion}
\end{equation}
with real coefficients $c_j$, and  $c_{rs}$ forming a complex-valued positive semidefinite matrix. The locality of the Lindbladian restricts the pairs of non-zero elements of $c_{rs}$; for instance, if the jump operators $L_j$ are on-site, $c_{rs}$ vanishes whenever $l_r, l_s$ act on different sites. Our goal is to infer the values of the non-zero coefficients $c_j$, $c_{rs}$.

To this end, we identify a set of local constraints that apply to any steady state $\rho_s$ of $\mathcal{L}$. Since $\rho_s$ is a steady state, the expectation value $\av{A}\EqDef \Tr\left({A\rho_s}\right)$ of any observable must be time-independent, 
\begin{equation}
\Tr\left({A\dot{\rho_s}}\right) = 0.
\end{equation}
Plugging in \Eq{lind_expansion} and using the cyclic properties of the trace, $\Tr\left({ABC}\right) = \Tr\left({CAB}\right)$ and $\Tr\left({A\com{B,C}}\right) = \Tr\left({C\com{A,B}}\right)$, we obtain the linear constraint
\begin{equation}
-\sum_i c_i \av{\i\com{A, h_i}} + \sum_{r,s} \frac{c_{rs}}{2} \av{\com{l_r, A}l_s^{\dagger} +  l_r \com{A, l_s^{\dagger}}} = 0,
\label{constraint}
\end{equation}
where the expectation values are taken with respect to the steady state $\rho_s$. For any operator $A$, \Eq{constraint} yields a linear equation for the parameters $c_j$ and $c_{rs}$. We will use a set of constraint operators $\lbrace{A_n\rbrace}$ to obtain a system of linear equations for the Lindbladian coefficients.

Importantly, assuming that local $A_n$ operators are chosen, the constraints derived from \Eq{constraint} are local in two ways. First, these constraints involve only local observables, which are easier to measure in most experimental settings. Second, if the $A_n$  operators act only within a given region, they commute with all the Lindblad terms that are supported outside that region. This allows to recover the Lindbladian of a region from measurements of that region alone.

We now introduce a convenient notation for representing the constraints derived from \Eq{constraint}. We concatenate the Hamiltonian parameters $c_j$ and the dissipative parameters $c_{rs}$ into a single vector $\vec{c}$. In this notation, \Eq{constraint} takes the form
\begin{equation}
\vec{k}^T \vec{c} = 0
\label{constraint_vector}
\end{equation} 
for a corresponding vector of expectation values $\vec{k}$. Since $c_{rs}$ is Hermitian, its upper and lower parts are redundant; each pair of off-diagonal elements contributes only a single pair of real parameters, $\Re{c_{rs}} = \frac{1}{2} \left( c_{rs} + c_{sr} \right)$ and $\Im{c_{rs}} = \frac{1}{2\i} \left(c_{rs} - c_{sr}\right)$. Thus, $\vec{c}$ is a real vector with four types of elements: Hamiltonian coefficients $c_j$, diagonal dissipative coefficients $c_{r,r}$, and the real and imaginary parts of the off-diagonal dissipative coefficients $c_{rs}$ for $r>s$. 

Repeating this procedure for a set of constraints $\lbrace{ A_n \rbrace}_{n=1}^{N}$, we obtain a homogeneous system of linear equations for the coefficients of the true Lindbladian,
\begin{equation}
K \vec{c} = 0,
\label{constraint_matrix}
\end{equation}
where $K$ is an ${N\times M}$ matrix of expectation values (see \App{app:exact_K}), with $N$ the number of constraints and $M$ the number of unknown parameters. Each of its rows corresponds to a constraint operator $A_n$, and each column to a different Hamiltonian term or jump operator appearing in \Eq{lind_expansion}. 

Assuming that we measured $K$ at a steady state of a local Lindbladian, the vector $\vec{c}$ corresponding to that Lindbladian must lie in the  kernel of $K$. If the steady state is shared by a family of Lindbladians, the kernel will be spanned by the whole family (see \App{app:fully_mixed} for the example of the fully mixed state). If the steady state corresponds to a unique local Lindbladian, the kernel of $K$ will become one-dimensional once sufficiently many constraints are used. We expect this to occur when the number of equations reaches the number of unknowns, revealing the true Lindbladian parameters up to an overall multiplicative constant. When a Lindbladian has multiple steady states, any of them may be used for the reconstruction; however, the reconstruction quality may depend on the steady state used. 

Thus, if the elements of $K$ are known exactly, our method recovers a unique Lindbladian whenever the equation
$K\vec{c}=0$ has a unique solution. Put differently, the
spectrum of singular values of $K$ must contain a single zero. In practice, the elements of $K$ are only known to a finite precision due to measurement noise. The spectrum of $K$ determines
the difficulty, or noise sensitivity, of the Lindbladian
reconstruction. 

Suppose that each observable is only measured to an additive error $\epsilon>0$ \footnote{For example, if each observable is measured experimentally using $n_s$ copies of $\rho_s$, its expectation value is known up to random noise of order $\epsilon \sim 1/\sqrt{n_s}$}. For the measured $K$, the equation $K \vec{c} = 0$ will likely not have an exact solution. As an approximate solution, we take the normalized coefficient vector $\hat{c} = \vec{c}/\norm{\vec{c}}$ that minimizes $\norm{K \hat{c}}$, i.e. the eigenvector of $K^T K$ with smallest eigenvalue. Since the Lindbladian is only recovered up to a multiplicative scalar, we measure the reconstruction error $\delta$ by the $L_2$ distance between the normalized coefficient vectors $\hat{c}$ of the recovered Lindbladian and the true Lindbladian,
\begin{equation}
\Delta = \norm{\hat{c}_{recovered} - \hat{c}_{true}}_2.
\label{error}
\end{equation}
Using perturbation theory, we estimated in \iRef{Bairey2019} the reconstruction error due to independent random noise with standard deviation $\epsilon$ added to each element of $K$,
\begin{equation}
\Delta^{est} = \epsilon \sqrt{\sum_{m>0} \lambda_{m}^{-1}},
\label{estimate}
\end{equation}
where $\lambda_m$ are the eigenvalues of $K^T K$ \footnote{In
particular, the reconstruction error is dominated by the gap $\lambda_1$
of the constraint matrix, since $\Delta_{est} \leq \epsilon \sqrt{M
\lambda_{1}^{-1}}$} (i.e., the squared singular values of $K$). 

\begin{figure*}  \includegraphics[width=\textwidth]{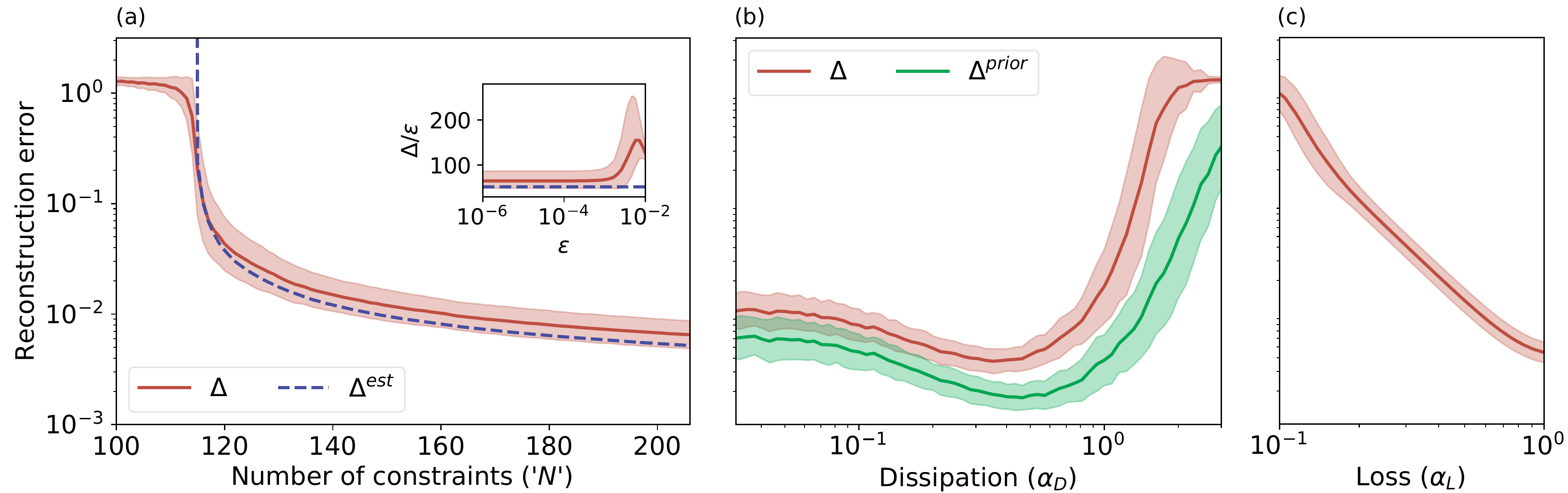} 
  \protect
  \caption{Reconstruction of  Lindbladians from their steady states. We generated steady states of random local Lindbladians on chains of $\Lambda=6$ spins and measured local observables given by \Eq{constraint} for a set of constraint operators $\lbrace{A_n\rbrace}_{n=1}^{N}$. We then recovered the Lindbladians from these observables by solving \Eq{constraint_matrix}, adding a small random measurement noise of order $\epsilon=10^{-4}$ to each observable, and computed the error $\Delta$ in the recovered Lindbladians [see \Eq{error}]. (a) Reconstruction error ($\Delta$) of random Lindbladians [Eqs. (\ref{hamiltonian}, \ref{dissipation})] as a function of the number of constraints (red; shaded area indicates error bars). Recovery succeeded once the number of constraints $N$ approached the number of unknowns $M$ (here $M=117$); its accuracy improved as more constraints were added, following the estimate $\Delta^{est}$ (dashed curve) of \Eq{estimate}. Here, the ratio between the magnitudes of the Hamiltonian and dissipation was fixed to $\alpha_D = \frac{1}{\sqrt{2}}$. Inset: the error-to-noise ratio $\Delta/\epsilon$ with all 3-local constraints as a function of the measurement noise magnitude $\epsilon$. The error followed the prediction of \Eq{estimate} as long as $\Delta \ll 1$. (b) Reconstruction error as a function of the dissipation strength $\alpha_D$. Here we used all constraints $A_n$ acting on up to 3 consecutive sites. Addition of weak dissipation improved the Lindbladian recovery, which was optimal at $\alpha_D \approx 0.5$. A lower reconstruction error was achieved when the Hamiltonian was known (green; $\Delta^{prior}$). (c) Dependence of the reconstruction error on the type of dissipation. We used the same ensemble of random Hamiltonians, with dissipation given by \Eq{dissoherence}, and $\alpha_L$ interpolating between loss and dephasing. When dissipation is almost entirely due to dephasing, $\alpha_L\to 0$, the steady state is close to being fully mixed; consequently, recovery improves with increasing loss (increasing $\alpha_L$). All results were averaged over 300 random Lindbladians, with error bars indicating one standard deviation; means and standard deviations were calculated after taking the log.} 
 \label{fig1}
\end{figure*}

\section{Results}

\subsection{Recovery of random local Lindbladians} We apply our method for the reconstruction of random local Lindbladians from their respective steady states. We start by focusing on chains of $\size=6$ spins with random local interactions and dissipation. We consider Lindbladians of the form given in Eq.(\ref{lind_def}) with local Hamiltonian terms
\begin{equation}\label{hamiltonian}
H_j = \sum_{\alpha=1}^{3} c_{j,\alpha}\sigma_j^\alpha + \sum_{\alpha,\beta=1}^{3} c_{j,\alpha,\beta} \sigma_j^\alpha \sigma_{j+1}^\beta,
\end{equation}
and on-site jump operators $L_j$ given by 
\begin{equation}\label{dissipation}
L_j = \sum_{\alpha=1}^{3} d_{j,\alpha}\sigma_j^\alpha.
\end{equation}
We choose open boundary conditions $c_{L,\alpha,\beta}=0$, and draw the remaining Hamiltonian coefficients from a Gaussian distribution with zero mean and unit variance, setting the energy scale for what follows. The real and imaginary parts of the dissipative coefficients $d_{j,\alpha}$ are similarly drawn from a Gaussian distribution, with mean zero and standard deviation $\alpha_D=\frac{1}{\sqrt{2}}$.

We obtain the steady state of each random Lindbladian $\mathcal{L}$ by exactly diagonalizing it as a superoperator. We then attempt to recover $\mathcal{L}$ using an increasing number $N$ of constraints $A_n$. We start with all the constraints $A_n$ acting on single sites and nearest neighbors, and add constraints supported on three consecutive sites in random order. To assess the reconstruction difficulty in practical settings, we add to each measured observable a small, independent, Gaussian noise with mean zero and standard deviation $\epsilon=10^{-4}$. We then compute the reconstruction error $\Delta$ due to the measurement noise $\epsilon$. 

As soon as the number of constraints approaches the number of unknowns, the reconstruction error $\Delta$ drops, and we obtain a good approximation of the Lindbladian (\Fig{fig1}a). The error decreases with the number of constraints, following the estimate of \Eq{estimate}. We verified numerically that the reconstruction error $\Delta$ follows the estimate of \Eq{estimate} over several orders of
magnitudes of the measurement noise $\epsilon$, as long as $\Delta \lesssim 10^{-2}$ (\Fig{fig1}a inset).

\subsection{Effect of dissipation type and strength} Next, we study how the accuracy of the method depends on the type and strength of the dissipative terms appearing in the Lindbladian. 
First, we vary the magnitude $\alpha_D$ of the dissipative terms appearig in Eq.~(\ref{dissipation}) relative to the Hamiltonian terms. We repeat the recovery experiment on the steady states of these different dynamics, using all 3-local constraints $A_n$. We find that the accuracy of the method improves upon adding weak dissipation to a Hamiltonian; the recovery is optimal when the dissipative terms are comparable in magnitude to the Hamiltonian terms (\Fig{fig1}b, red). Due to our choice of single-site jump operators in Eq.~(\ref{dissipation}),  steady states at the strong dissipation limit approach product states. Since any product state is a steady state of many different Lindbladians, the reconstruction error diverges for $\alpha_D \rightarrow \infty$; this divergence of the error is cured when two-site nearest-neighbor jump operators are added (see \App{app:strong_dissipation}).

In practical situations, the jump operators $L_j$ may be unknown even if the Hamiltonian is well-characterized. We can incorporate prior knowledge about the Hamiltonian by turning \Eq{constraint} into the non-homogeneous constraint
\begin{equation}
\sum_{r,s} \frac{c_{rs}}{2} \av{\com{l_r, A}l_s^{\dagger} +  l_r \com{A, l_s^{\dagger}}} = \av{\i\com{A, H}},
\label{constraint_non_homogeneous}
\end{equation}
where the RHS is directly obtained by measurements. The dissipative coefficients $c_{rs}$ are then obtained by solving a system of non-homogeneous linear equations (see \App{app:prior}). \Fig{fig1}b shows that recovery with such prior knowledge of the Hamiltonian 
achieves a lower reconstruction error of the Lindbladian (green curve). Since the recovery with prior knowledge leaves no ambiguity in the magnitude of the Lindbladian, we can also compare the dynamics generated by the true and recovered Lindbladians starting from a fixed initial state; indeed, we find an excellent agreement (\Fig{reconstructed_dynamics}).

Next, we study the interplay of different dissipation types. We consider a Lindbladian $\mathcal{L}$ which consists of single-site jump operators of two kinds:
\begin{equation}
L_{j,L}=\alpha_L \sigma^-_j, \hspace{0.5cm}  L_{j,D}=(1 - \alpha_L) \sigma^z_j,
\label{dissoherence}
\end{equation} 
where $\sigma^- \EqDef \frac{1}{2}\left(\sigma^x - \i \sigma ^y\right)$. The ``loss" $L_{j,L}$ relaxes the system towards a pure steady state, e.g. due to loss of particles; the ``dephasing" $L_{j,D}$ scrambles relative phases between pure states in a specific basis. We tune the parameter $0\leq \alpha_L\leq 1$ to interpolate the relative weights of the loss and dephasing. In addition, $\mathcal{L}$ contains Hamiltonian terms of the form (\ref{hamiltonian}), with coefficients drawn from a Gaussian distribution with zero mean and unit variance. We then attempt to recover both the Hamiltonian and the jump operators from the steady state of $\mathcal{L}$ using all 3-local constraints $A_n$, without assuming that the form of the on-site jump operators is known.

We find that reconstruction of strongly dephasing Lindbladians is hard (\Fig{fig1}c). This is expected: for $\alpha_L \ll 1$, the steady state is close to a fully mixed state, compatible with any Lindbladian with Hermitian jump operators. As the loss intensifies, $\norm{\mathcal{L}(\mathbbm{1})} \propto \alpha_L^2$; correspondingly, \Fig{fig1}c shows that the reconstruction error decreases as $\alpha_L^{-2}$ (see also \App{app:loss_to_dephasing}), indicating that the steady state becomes more informative.

\begin{figure}[t]
\includegraphics[width=8.6cm]{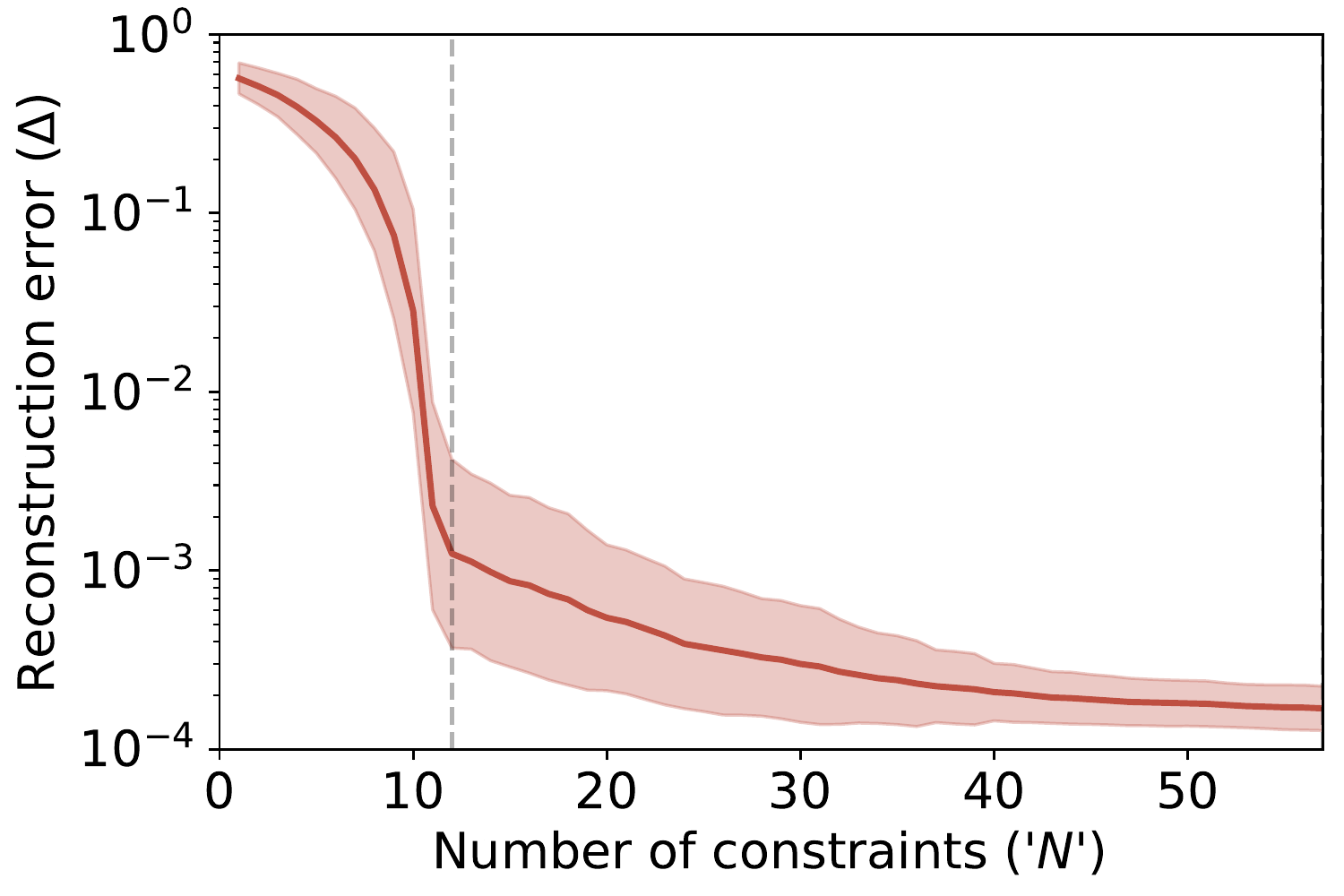} 
\protect\caption{Loss facilitates learning of commuting Hamiltonians: error in the reconstruction of classical Hamiltonians from steady states of dissipative dynamics, as a function of the number of constraints $N$. We generate random classical Ising Hamiltonians on a one-dimensional chain with $\Lambda=6$ spins [\Eq{ising}]. While these Hamiltonians are impossible to learn from a generic steady state, the addition of loss $L_j = 2 \sigma_j^-$ allows to extract their coupling parameters. Due to the small number of unknowns (only $M=11$ Hamiltonian terms), recovery is easy, and single-site constraint operators suffice (dashed vertical line; corresponds to 2-local measured observables). } \label{fig2}
\end{figure}

\subsection{Loss facilitates learning of commuting Hamiltonians} Motivated by the insight that loss can lead to non-trivial steady states, we investigate whether dissipation can aid in learning Hamiltonians that could not be recovered from their own steady states. In particular, we consider classical Hamiltonians with random nearest-neighbor interactions in the X-basis alone,
\begin{equation}
H_{cl}^{x} = \sum_{j=1}^{\Lambda} b_j \sigma_j^x + \sum_{j=1}^{\Lambda-1} J_j \sigma_j^x \sigma_{i+1}^x,
\label{ising}
\end{equation}
whose coefficients are drawn from a Gaussian distribution with zero mean and unit variance. Any state $\rho$ diagonal in the X-basis is a steady state of $H_{cl}^{x}$, revealing no information about its coefficients. We therefore add on-site jump operators 
\begin{equation}
L_j = 2 \sigma_j ^- ,
\end{equation} 
so that the dynamics of $\mathcal{L}$ are comprised of Hamiltonian dynamics in the $X$ basis and loss in the $Z$ basis. We then attempt to recover $H$ from the steady state of $\mathcal{L}$, assuming that the jump operators $L_j$ are known. 

We find that the addition of controlled loss facilitates efficient learning of the classical Hamiltonians of \Eq{ising}. Due to the small number of unknowns, single-site constraint operators $\sigma_j^y$, $\sigma_j^z$ are sufficient to recover $H$ ($\sigma_j^x$ are not required as they commute with $H$). Moreover, the reconstruction is very robust: when nearest-neighbor constraints are added, the accuracy of the recovered Hamiltonian approaches the measurement accuracy (\Fig{fig2}).

\begin{figure}[t]
\includegraphics[width=8.6cm]{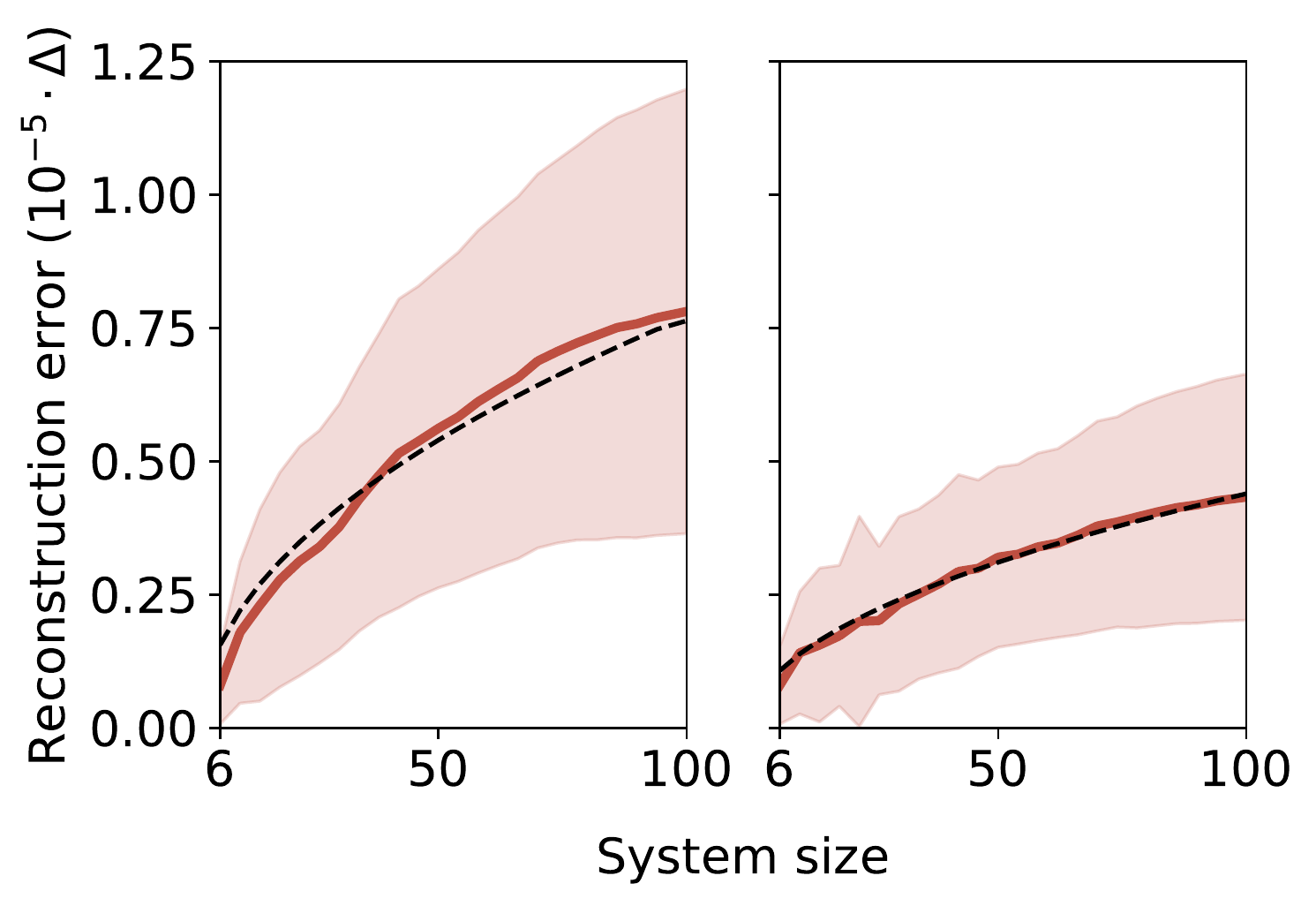} 
\protect\caption{Reconstruction of Lindbladians on large spin chains: system-size scaling of the reconstruction error. We obtained the steady states of the random Lindbladians described in Eqs. (\ref{hamiltonian}), (\ref{dissipation}) on $\Lambda=100$ spins. (left) We recovered the Lindbladians on spatial patches of $6$ spins, with overlaps of $2$ sites between consecutive patches. We used all constraints supported on up to $3$ consecutive sites in the interior of each patch (middle $4$ sites for bulk patches). We then stitched consecutive patches to obtain the full Lindbladian on subsystems of increasing length. The reconstruction error increased with system size (red curve), following the predicted square-root scaling with the number of patches (dashed curve). (right) As a different approach, we built a single large constraint matrix for each subsystem, and obtained the error as a function of subsystem size; this approach yielded a slightly smaller reconstruction error, still scaling as the square root of subsystem size (dashed curve). } \label{fig3}
\end{figure}
\subsection{System-size scaling}
Finally, we demonstrate that our method can recover Lindbladians on long spin chains. Various approaches have been proposed for computing steady states of large-scale open quantum systems using matrix product operators ~\cite{Cui2015, Mascarenhas2015, Werner2016}. In this work, we have used the variational MPO approach of ~\iRef{Mascarenhas2015}, which iteratively finds the density matrix with the smallest-magnitude eigenvalue of $\mathcal{L}$. Using this approach, we obtain steady states of the random Lindbladians considered in Eqs. (\ref{hamiltonian}), (\ref{dissipation}) on chains with $\size=100$ spins (see \App{app:MPO} for details). 

To study the system-size scaling of our method, we focus on subsystems of increasing sizes. We begin with the $6$ leftmost spins and add $4$ spins in each step, eventually covering the whole chain. We then attempt to recover the Lindbladian of each of these subsystems from observables within that subsystem only, using all 3-local constraints. 

We employ two different approaches for recovering the full Lindbladians of these increasingly large subsystems. In the first approach, we partition the subsystem to overlapping patches of $6$ spins, and recover the Lindbladian on each patch independently. The recovery does not determine the overall scale factor of the Lindbladian on the patch; we therefore re-scale the coefficients of neighboring patches according to the coefficients of their shared terms (see \App{app:scaling}). In the second approach, we apply our method directly on the whole subsystem, forming a large constraint matrix $K$ which grows with the subsystem size. 

Both approaches successfully recover the full-system Lindbladian using the same set of measurements. Here we do not add measurement noise; the error in a single patch ($\approx 10^{-6}$) is controlled by the numerical precision of the MPO steady state. Due to the uncertainty in the coefficients shared between each pair of patches, the norm of the recovered Lindbladian performs a random walk, leading to a total error growing as the square root of the number of patches (\Fig{fig3}, left; see \App{app:scaling} for analysis). Namely, the error grows as the square root of system size, $O(\Lambda^{\frac{1}{2}})$. We find the same square root system-size scaling of the reconstruction error in the second, direct approach (\Fig{fig3}, right).

These findings suggest that in order to recover the dynamics of a system of length $\Lambda$ to a fixed accuracy, each observable should be measured to an accuracy of $O(\Lambda ^{-\frac{1}{2}})$. In other words, each observable should be measured $n_s = O(\Lambda)$ times. The number of observables required scales also as $O(\Lambda)$; however, since they are all local, each copy of the steady state $\rho_s$ can be used to measure $O(\Lambda)$ observables. Thus, we expect that $n_s = O(\Lambda)$ copies of $\rho_s$ overall suffice.

\section{Conclusions} Near-term intermediate-scale quantum devices \cite{Preskill2018} are invariably subject to noise and coupled to their environments. While tomographic methods can characterize noises acting on a few isolated qubits \cite{Merkel2013, Blume-Kohout2017}, cross-talk between qubits necessitates holistic methods that identify the sources of error in an entire device \cite{Proctor2018}. 

Our results suggest that the noises acting on quantum devices may be efficiently characterized from measurements of their steady states. Left to themselves, quantum devices naturally reach their steady states at times longer than their typical relaxation and decoherence timescales. If in addition to single-qubit dissipation, the qubits are also  coupled by a Hamiltonian or affected by correlated dissipation, we find that their steady state would be informative enough to recover both the Hamiltonian and the dissipative processes.

In addition to scalability, our approach to characterizing dynamics through their steady states offers a few advantages. It does not require precise control of either state initialization or measurement time. It is independent on the dimensionality of the local Hilbert space, and is effective also for bosonic systems with an infinite-dimensional local Hilbert space. As shown in \Fig{fig2}, addition of controlled terms can allow learning of Hamiltonians consisting of commuting terms, such as those corresponding to topological quantum error-correcting codes \cite{Valenti2019}.

Having demonstrated that open quantum system dynamics can generically be learned from their steady states, it is important to obtain rigorous bounds on the number of measurements required for the learning process. Such bounds could be obtained by identifying conditions under which our constraint matrix is guaranteed to be gapped. It could also be interesting to study our method as a means to certify quantum states prepared as the steady states of given quantum dynamics. Finally, adapting our method to the setting of quantum circuits may yield means to certify, characterize and benchmark quantum devices.

\begin{acknowledgments}
We thank Yotam Shapira for useful discussions.  E. B. and N. L. acknowledge financial support from the
 European Research Council (ERC) under the European Union Horizon 2020 Research and Innovation Programme (Grant Agreement No. 639172). D.P. acknowledges support from the Singapore Ministry of Education, Singapore Academic Research Fund Tier-II (project MOE2018-T2-2-142).  N. L. acknowledges support from the People Programme  (Marie Curie Actions) of the European Union\textquoteright s Seventh Framework Programme (No. FP7/2007\textendash 2013) under  REA Grant Agreement No. 631696 and the Defense Advanced Research Projects Agency through the DRINQS program, grant No. D18AC00025. The content of the information presented here does not necessarily reflect the position or the policy of the U.S. government, and no official endorsement should be inferred. I.A.  acknowledges the support of the Israel Science Foundation (ISF) under the Individual Research Grant No. 1778/17.
\end{acknowledgments}

\bibliographystyle{apsrev4-1}

\bibliography{library}


\renewcommand{\theequation}{S\arabic{equation}}
\renewcommand{\thefigure}{S\arabic{figure}}
\setcounter{equation}{0}
\setcounter{figure}{0}
\appendix

\section{Details of the recovery algorithm}

\subsection{Expanding the Lindblad dynamics in a fixed set of operators: derivation of \Eq{lind_expansion}} \label{app:expansion}
Formally, to derive \Eq{lind_expansion} from \Eq{lind_def}, we first expand each local Hamiltonian term in a fixed basis of local operators
\begin{equation}
    H_j = \sum_i c^{(j)}_i h_i,
\end{equation}
so that the unitary evolution term becomes
\begin{equation}
    \sum_j \com{H_j, \rho} = \sum_i c_i \com{h_i, \rho}
\end{equation}
with
\begin{equation}
    c_i = \sum_j c^{(j)}_i.
\end{equation}

Similarly, we expand each jump operator in a fixed basis of local operators
\begin{equation}
    L_j = \sum_i c^{(j)}_r l_r,
    \label{jump_expansion}
\end{equation}
so that the dissipative dynamics may be rewritten as 
\begin{align*}
    \frac{1}{2} \sum_j \left(\com{L_j \rho, L_j^{\dagger}} + \com{L_j, \rho L_j^\dagger} \right) = \\
    = \sum_{r,s} \frac{c_{rs}}{2} \left( \com{l_r \rho, l_s^{\dagger}} + \com{l_r, \rho l_s^{\dagger}} \right),
\end{align*}
where
\begin{equation}
    c_{rs} = \sum_{j} c_r^{(j)} (c_s^{(j)})^*
\end{equation}
forms a positive semi-definite matrix by definition.

\subsection{Exact form of the constraint matrix} \label{app:exact_K}

\begin{figure}
  \includegraphics[width=8.6cm]{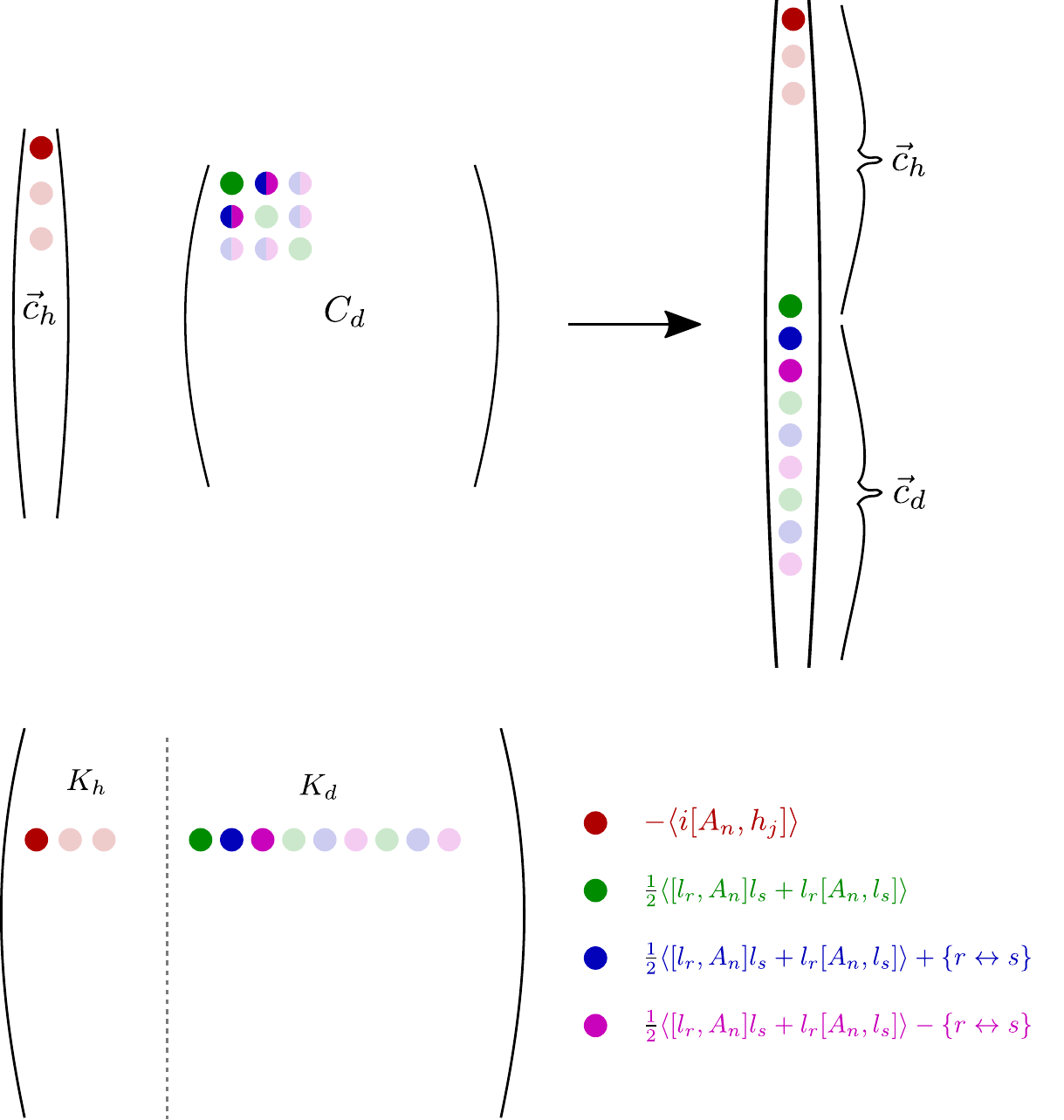} 
  \protect
  \caption{Top: we concatenate the Hamiltonian coefficients $\vec{c}_h$ and the matrix of dissipative coefficients $C_d$ into a long vector of coefficients for the Lindblad evolution. Off-diagonal entries of $C_d$ are split into their real and imaginary parts (blue and magenta, correspondingly). Bottom: the constraint matrix is composed of a vertical block corresponding to Hamiltonian terms $K_h$, and a vertical block corresponding to dissipative terms $K_d$. Entries corresponding to Hamiltonian terms are given by their commutators with the constraint operators (red); the formula for the dissipative entries varies between the diagonal entries of the dissipative matrix $C_d$ (green), and the real (blue) and imaginary (magenta) entries of $C_d$.} 
 \label{K_fig}
\end{figure}

As derived in Eqs (\ref{constraint}-\ref{constraint_matrix}), the elements of the constraint matrix $K$ are expectation values of different observables. The explicit form of the element $K_{n,m}$ varies, depending on the term in the expansion of the Lindbladian in \Eq{lind_expansion} which corresponds to the index $m$: (i) coefficients $c_j$ of Hamiltonian terms; (ii) diagonal entries of the matrix of dissipative coefficients $c_{rr}$; (iii) the real part of the off-diagonal dissipative coefficients $\frac{1}{2}\Re{c_{rs} + c_{sr}}$; (iiii) the imaginary part of the off-diagonal dissipative coefficients $\frac{1}{2\i}\Re{c_{rs} - c_{sr}}$. Explicitly, the matrix elements $K_{n,m}$ are given by (see also \Fig{K_fig}):

\begin{equation}
\begin{aligned}
& K_{n,m} = \\
& 
\begin{cases}
-\av{i\com{A_n,h_j}} & c_j \\
\frac{1}{2} \av{\com{l_r, A_n}l_s^{\dagger} +  l_r \com{A_n, l_s^{\dagger}}} & c_{rr} \\
\frac{1}{2} \left( \av{\com{l_r, A_n}l_s^{\dagger} +  l_r \com{A_n, l_s^{\dagger}}} + \left\lbrace r \leftrightarrow s \right\rbrace \right) & \Re{c_{rs}}; r > s \\
\frac{i}{2} \left( \av{\com{l_r, A_n}l_s^{\dagger} +  l_r \com{A_n, l_s^{\dagger}}} - \left\lbrace r \leftrightarrow s \right\rbrace \right) & \Im{c_{r,s}}; r > s. \\
\end{cases}
\end{aligned}
\end{equation}

\subsection{\Eq{constraint} for a fully mixed state} \label{app:fully_mixed}

From \Eq{lind_def} it is clear that the fully mixed steady state $\rho \propto \mathbbm{1}$ is a steady state of any Lindbladian with Hermitian jump operators $L_j = L_j^\dagger$ (in fact, it is sufficient that the jump operators are normal, $\com{L_j, L_j^\dagger}=0$). Let us see how this reflects in \Eq{constraint}.

If the dissipators $L_j$ are real, we can  expand them (see \Eq{jump_expansion}) in a basis of Hermitian local operators $l_r = l_r^{\dagger}$ using real coefficients; subsequently, the coefficient matrix $c_{rs}$ will be real and symmetric. At a fully mixed state, the expectation value of any operator is proportional to its trace, and \Eq{constraint} becomes
\begin{equation}
-\sum_i c_i \Tr \left( \i\com{A, h_i} \right) + \sum_{r,s} \frac{c_{rs}}{2} \Tr \left( \com{l_r, A}l_s +  l_r \com{A, l_s} \right) = 0.
\label{constraint_fully_mixed}
\end{equation}
Since commutators are traceless $\Tr \com{A,B} = \Tr AB - \Tr BA = 0$, the first part vanishes; in other words, the fully mixed state is a steady state of any Hamiltonian. We now note that the second term is antisymmetric in $r \leftrightarrow s$: using the cyclic properties of the trace $\Tr AB = \Tr BA$ and $\Tr \left( A \com{B,C} \right) = \Tr \left( C \com{A,B} \right) $,
\begin{align}
    & \Tr \left( \com{l_r, A}l_s \right) +  \Tr \left(l_r \com{A, l_s} \right) \\
    &= \Tr \left( \com{l_s, l_r} A \right) +  \Tr \left(A \com{l_s, l_r} \right) \\ 
    &= 2 \Tr \left( \com{l_s, l_r} A \right),
\end{align}
which is antisymmmetric to $r \leftrightarrow s$ due to the commutator. On the other hand,  $c_{rs}$ is symmetric, so the sum over $r,s$ vanishes:
\begin{equation}
    \sum_{r,s} c_{rs} \Tr \left( \com{l_s, l_r} A \right) = 0.
\end{equation}
Thus, the fully mixed state obeys \Eq{constraint} for any constraint operator $A$ if the jump operators $L_j$ are Hermitian.

\subsection{Recovery with prior knowledge} \label{app:prior}
If some part of the dynamics is known to high accuracy, \Eq{constraint} can be turned into a non-homogenous equation. For instance, if the Hamiltonian is known but the dissipators are not, we obtain \Eq{constraint_non_homogeneous}. Using a set of constraint operators $\lbrace{A_n \rbrace}_{n=1}^{N}$, we obtain the system of equations
\begin{equation}
    K_l \vec{c_l} = \vec{b}, 
    \label{K_prior}
\end{equation}
where $K_l$ is the constraint matrix of the dissipative operators alone, and $c_l$ are their corresponding coefficients; the vector $\vec{b}$ is given by
\begin{equation}
    b_n = \av{\i \com{A_n, H}}.
\end{equation}
\Eq{K_prior} is then solved using least squares.

\section{Error analysis}

\subsection{Recovery of strongly dissipating Lindbladians} \label{app:strong_dissipation}
In \Fig{fig1}b, it appears that the recovery error diverges when the relative magnitude of the dissipative terms is large $\alpha_D > 1$. We conjectured that this divergence does not indicate that recovery is generically impossible in the limit of strong dissipation; rather, it is an artifact of the choice of strictly single-site dissipation we simulated.

\begin{figure}
  \includegraphics[width=8.6cm]{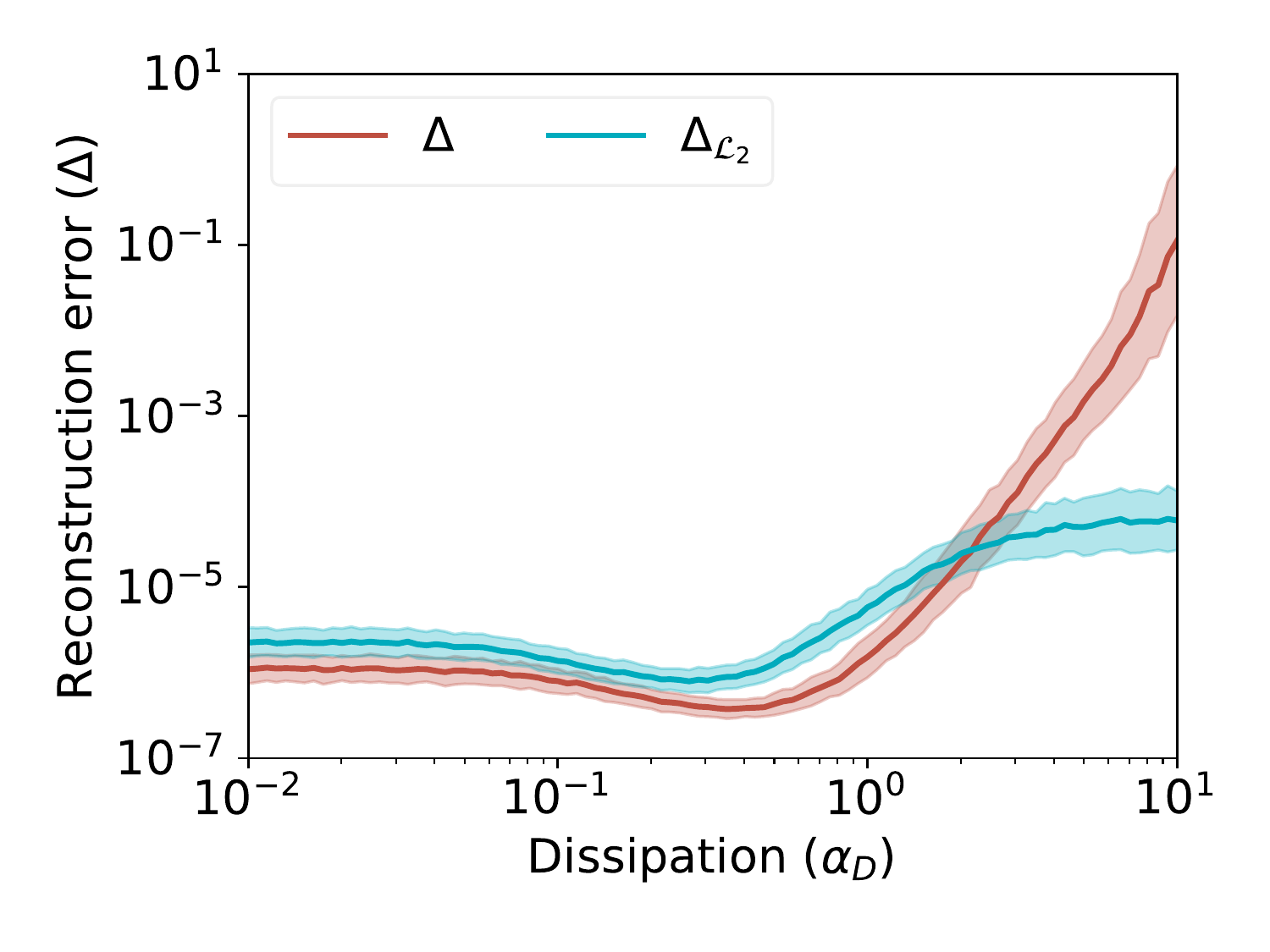} 
  \protect
  \caption{Entangling jump operators facilitate learning of strongly dissipative dynamics. Reconstruction error as a function of dissipation strength $\alpha_D$ of Lindbladians with  nearest-neighbor Hamiltonian terms [\Eq{hamiltonian}]; for the dissipation, we took either strictly single-site jump operators (red), as in main text [\Eq{dissipation}]; or both single-site and  nearest-neighbor jump operators (blue) [\Eq{dissipation_nn}].  The divergence of the error at the strong dissipation limit is cured when nearest-neighbor jump operators are added. Here, we added a smaller noise than in the main text ($\epsilon = 10^{-8}$ rather than $10^{-4}$) to probe the behavior at large values of $\alpha_D$.} 
 \label{balance_supp}
\end{figure}

To verify this conjecture, we added nearest-neighbor jump operators to our random Lindbladians
\begin{equation}
    L_j = \sum_{\alpha=1}^{3} d_{j,\alpha} \sigma_j^{\alpha} + d_{j,x,x} \sigma_{j}^{x} \sigma_{j+1}^{x}+ d_{j,y,y} \sigma_{j}^{y} \sigma_{j+1}^{y},
    \label{dissipation_nn}
\end{equation}
with all coefficients drawn from a Gaussian distribution with mean zero and standard deviation $\alpha_D$; for the Hamiltonian terms, we used the same random nearest-neighbor interactions of \Eq{hamiltonian}. We then recovered these Lindbladians from their steady states, assuming that the form of the jump operators is known but their coefficients are not. We found that the reconstruction error of these Lindbladians saturates at large $\alpha_D$ (\Fig{balance_supp}, blue); thus, the divergence of the reconstruction error is cured when entangling jump operators are added.

\subsection{Recovery error: results vs. expectation} \label{app:results_vs_expectation}

The recovery error $\Delta$ we find in \Fig{fig1}a is slightly higher (by a factor of $\approx 1.25$) than the estimate of \Eq{estimate}, derived in \iRef{Bairey2019}. In contrast to our results in this work, the recovery error obtained in \iRef{Bairey2019} was lower than the prediction of the same estimate, which is indeed expected to be pessimistic due to the use of Jensen's inequality.  

We believe the difference is due to the different noise model used in both papers: here we add noise to each measured observable, while in \iRef{Bairey2019} we added independent noise to each of the entries of $K$ (even when they contain the same observable). This is because in \iRef{Bairey2019}, we wished to test the theoretical validity of the error estimate. The estimate assumes that the noise in each entry of the constraint matrix $K$ is independent, and we thus added an independent random noise to each of its entries. Realistically though, noise is incurred in each measured observable. Since many different entries of $K$ feature the same observable, this introduces correlations between the noise in different entries. 

\subsection{Accuracy of the reconstructed dynamics} \label{app:dynamics}

To assess how well the recovered dynamics approximate the true dynamics, we compared the time evolution generated by the recovered and true Lindbladians starting from a fixed initial state. We focused on random Lindbladians with a relative dissipation magnitude $\alpha_D = \frac{1}{\sqrt{2}}$ and a known Hamiltonian, exactly as in \Fig{fig1}b (green curve). The knowledge of the Hamiltonian allows to recover the Lindbladian exactly (including its overall magnitude), allowing a meaningful comparison of time dynamics.

We initialized the system in a product state with all spins up,
\begin{equation}
    \rho(0) = \ketbra{\uparrow \uparrow \cdots \uparrow}{\uparrow \uparrow \cdots \uparrow},
\end{equation}
and computed its evolution under the true Lindbladian $\rho (t)$ and under the recovered Lindbladian $\rho_{rec}(t)$. At each point in time, we compared these two states by the average trace distance between their reduced density matrices on pairs of consecutive sites,
\begin{equation}
    D_{loc} (\rho, \rho_{rec}) = \frac{1}{\Lambda - 1} \sum_{i=1}^{\Lambda-1} D \left(\rho^{(i,i+1)}, \rho_{rec}^{(i,i+1)} \right),
\label{D_loc}
\end{equation}
where $\rho^{(i,j)} = \Tr_{\Lambda \setminus \lbrace{i,j \rbrace}} \left( \rho \right)$ is the reduced density matrix on sites $i,j$, and the trace distance
\begin{equation}
    D(\rho, \sigma) = \frac{1}{2} \norm{\rho - \sigma}_1
\end{equation}
bounds the difference in the expectation value of any POVM element. Thus, $D_{loc} (\rho, \rho_{rec})$ is a worst-case measure for the difference between local observables in the two states. 

\begin{figure}
  \includegraphics[width=8.6cm]{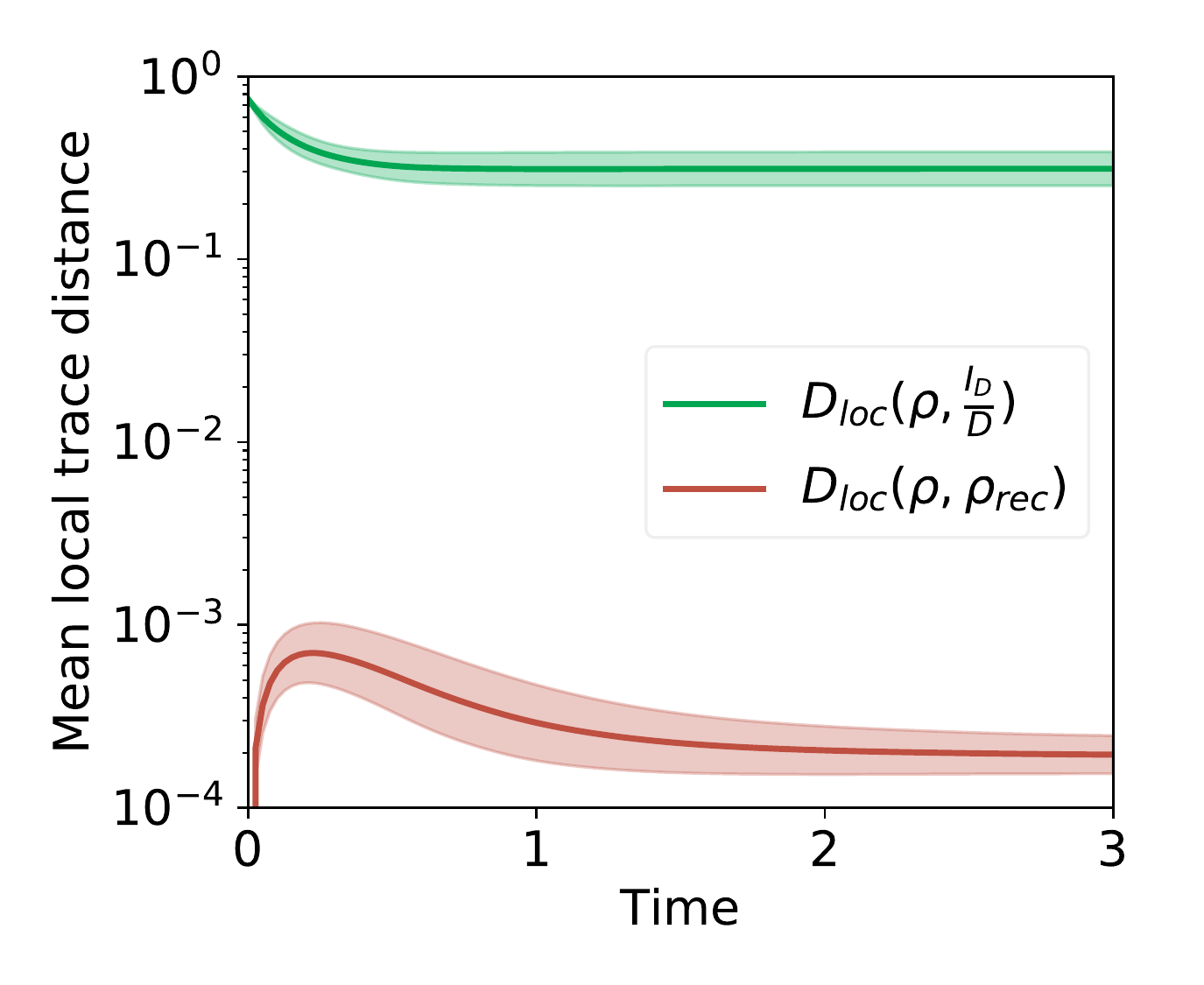} 
  \protect
  \caption{The accuracy of the evolution generated by the reconstructed dynamics as a function of time. We initialize the system in a product state with all spins up. We then measure the deviation between its evolution by the true dynamics $\rho(t)$ and its evolution by the recovered dynamics $\rho_{rec}$ by the mean local trace distance (red curve, see \Eq{D_loc}). For comparison, we also show the mean local trace distance between the true dynamics and the fully mixed state (green curve).} 
 \label{reconstructed_dynamics}
\end{figure}

As shown in \Fig{reconstructed_dynamics}, the mean local trace distance peaks at a value below $10^{-3}$ for short times. It then decreases to $\approx 2 \cdot 10^{-4}$, which is approximately the measurement accuracy taken for the reconstruction. This is not surprising in retrospect: at long times, $\rho_{rec}$ is the steady state of the recovered Lindbladian, which was chosen such that the measured local observables would correspond to its steady state.

\subsection{Scaling of the reconstruction error with the relative weight of loss in the dissipation} \label{app:loss_to_dephasing}

\begin{figure}
  \includegraphics[width=8.6cm]{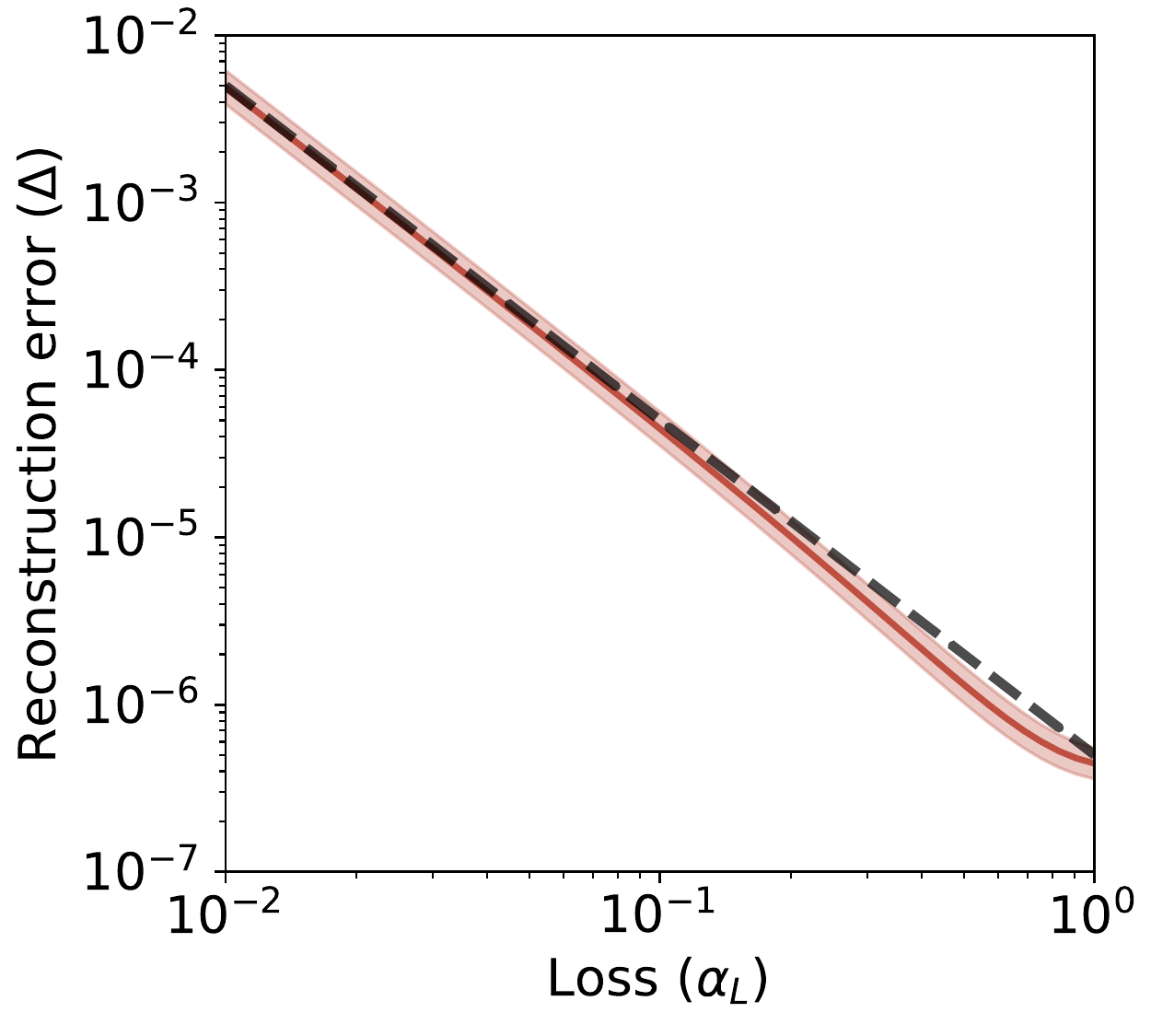} 
  \protect
  \caption{Reconstruction error as a function of the relative weight of loss in the dissipation $\alpha_L$. We repeated the simulations of \Fig{fig1}c with lower measurement noise $\epsilon=10^{-8}$ over a wider range of $\alpha_L$. The dashed line follows the equation $y=50 \epsilon / x^2$, confirming the theoretical expectation for the scaling of the reconstruction error with $\alpha_D$. } 
 \label{dissoherence_supp}
\end{figure}

We argued that \Fig{fig1}c confirms the theoretical expectation that the reconstruction error scale as $\alpha_L^{-2}$ when the weight of loss relative to dephasing $\alpha_L$ [see \Eq{dissoherence}] is small. However, the curve in \Fig{fig1}c did not show a clear power law for small $\alpha_L$, since the reconstruction error approached large values of order $1$. We thus repeated these simulations with weaker measurement noise ($\epsilon=10^{-8}$ compared to $\epsilon=10^{-4}$ in the main text), and verified this power law over a wider range of $\alpha_L$ (\Fig{dissoherence_supp}).

\section{Computing the steady state using variational matrix product operators} \label{app:MPO}
The steady state of the Lindbladian can be obtained by computing the eigenstate of the Lindblad operator $\mathcal{L}$ corresponding to eigenvalue $0$\cite{Cui2015, Mascarenhas2015} (the system we studied has no degeneracy). Internally the density operator $\rho$ is reorganized into a long vector and treated similarly to the state vector of a unitary system. We use the variational matrix product operator algorithm proposed in \iRef{Mascarenhas2015}, where the iterative procedure to search for the steady state is done in the same way as the unitary case, except that one keeps the eigenstate corresponding to the eigenvalue with smallest magnitude instead of smallest algebraic value. For a system with $100$ spins, we have used $D=100$ number of states (bond dimension) for our simulations. We obtained a steady state with eigenvalue of the order $10^{-8}$, and residual $\norm{\mathcal{L}\rho} \approx 10^{-5}$. To check the convergence against different $D$s, we have done another simulation with $D=150$, and compared the local observables $\langle\sigma^z_j\rangle$, obtaining a mean error $\sum_{j=1}^{\Lambda} \abs{\langle\sigma^z_j\rangle_{D=100}-\langle\sigma^z_j\rangle_{D=150}}/\Lambda \approx 10^{-9}$. We also compared the distances between the reduced density matrices with a patching size $6$, and a patching spacing $4$, and obtained a mean error of the order of $10^{-8}$. 

\section{Stitching up recovered patches} \label{app:scaling}

Recall that the Lindbladian on each patch is only recovered up to a multiplicative scalar. Suppose we recover the Lindbladian of two overlapping patches and wish to ``stitch'' them together into one Linbladian acting on the joint patch. In the absence of noise, the recovered Lindbladians of the first two patches would be given by
\begin{equation}
\begin{cases}
\vec{c}_l \cdot \mathcal{L}_{l} + \vec{c}_{m} \cdot \mathcal{L}_{m} \\ 
\pvec{c}'_{m} \cdot \mathcal{L}_{m} + \vec{c}_{r} \cdot \mathcal{L}_{r},
\end{cases}
\end{equation}
where $\mathcal{L}_{m}$ is the vector of terms [$h_j$ and pairs $(l_r, l_s)$] acting on the overlapping region of the two patches; for the analysis below, we assume that each individual recovered Lindbladian is normalized: $\norm{\vec{c}_l}^2 + \norm{\vec{c}_m}^2 = \norm{\pvec{c}'_m}^2 + \norm{\vec{c}_r}^2 = 1$. The coefficients $c_m, \pvec{c}'_m$ of the overlapping region will generically differ since the Lindbladian on each patch is only recovered up to a multiplicative scalar. We therefore use these overlapping coefficients to determine the relative scale of the two patches, by multiplying the Lindbladian of the second patch by a factor of $\frac{\norm{\vec{c}_m}}{\norm{\pvec{c}'_m}}$:
\begin{equation}
\mathcal{L}_{stitched} = \vec{c}_l \cdot \mathcal{L}_l + \vec{c}_m \cdot \mathcal{L}_m + \frac{\norm{\vec{c}_m}}{\norm{\pvec{c}'_m}} \vec{c}_r \cdot \mathcal{L}_r.
\end{equation}
In fact, we also need to fix the relative signs of the two patches using a similar factor of $\frac{\text{sign}({\vec{c}_m})}{\text{sign}({\pvec{c}'_m})}$, where the sign can be determined e.g. according to the coefficient of a fixed shared term. While this last detail is crucial for the stitching process, it does not contribute to the recovery error due to noise, as long as the error in each patch is small relative to its size, so that no coefficient flips its sign.

To recover the Lindbladian of a sequence of patches $1,\dots,n$, we repeat this procedure iteratively and obtain
\begin{equation}
\mathcal{L}_{stitched}^{(n)} = \sum_{j=1}^{n} \mathcal{L}_{patch}^{(n)},
\end{equation}
where
\begin{equation}
\mathcal{L}_{patch}^{(1)} = \vec{c}_1 \cdot \mathcal{L}_1 + \vec{c}_{1,2} \cdot \mathcal{L}_{1,2},
\end{equation}
with $\mathcal{L}_{1,2}$ denoting the terms acting on the overlapping region of the first two patches. For any $j>1$,
\begin{equation}
\mathcal{L}_{patch}^{(j)} = \left(\prod_{i=1}^{j-1} \frac{\norm{\vec{c}_{i,i+1}}}{\norm{\pvec{c}'_{i,i+1}}} \right)\left(\vec{c}_{j} \cdot \mathcal{L}_{j} + \vec{c}_{j,j+1} \cdot \mathcal{L}_{j,j+1} \right).
\end{equation}
If each individual patch is recovered perfectly up to a corresponding multiplicative scalar, this procedure yields the full system Lindbladian up to a single overall multiplicative scalar. However, noise introduces error in the recovered Lindbladian of each individual patch: $\vec{c}_j \mapsto \vec{c}_j + \vec{\delta}_j$. 

Error in each individual patch affects the overall stitched Lindbladian in two ways. One effect is a rotation of each $\mathcal{L}_{patch}$ with respect to its true value, the $\mathcal{L}_j$ component pointing to $\vec{c}_j + \vec{\delta}_j$ rather than $\vec{c}_j$. Since this error is additive, it is absorbed in the normalization of $\mathcal{L}_{stitched}$; assuming that the error is approximately uniform across patches, $\norm{\vec{\delta}_j} \approx \delta$, it leads to an overall error of order $\delta$ in the total $\mathcal{L}_{stitched}$, which is independent of the number of patches.

A second effect caused by the errors in the recovery of individual patches is a stretch of each $\mathcal{L}_{patch}$. This effect is induced through the errors' effect on the relative scale factor $\prod_{i=1}^{j-1} \frac{\norm{\vec{c}_{i,i+1}}}{\norm{\pvec{c}'_{i,i+1}}}$.  Assuming that the errors of the different patches $\vec{\delta}_j$ are independent, this scale factor performs a multiplicative random walk, fluctuating from its true value by a deviation of order $\sqrt{j} \delta$. This is most easily seen by taking a log:
\begin{align}
& \log \left( \prod_{i=1}^{j-1} \frac{\norm{\vec{c}_{i,i+1} + \vec{\delta}_{i,i+1}}}{\norm{\pvec{c}'_{i,i+1} + \pvec{\delta}'_{i,i+1}}}  \right) - \log \left( \prod_{i=1}^{j-1} \frac{\norm{\vec{c}_{i,i+1}}}{\norm{\pvec{c}'_{i,i+1}}}  \right)
\\
&= \sum_{i=1}^{j-1} \left( \log \frac{\norm{\vec{c}_{i,i+1} + \vec{\delta}_{i,i+1}}}{\norm{\vec{c}_{i,i+1}}} - \log \frac{\norm{\pvec{c}'_{i,i+1} + \pvec{\delta}'_{i,i+1}}}{\norm{\pvec{c}'_{i,i+1}}} \right).
\label{log_scale_factor}
\end{align}
To first order in $\delta$, each of these is an independent random variable with zero mean and standard deviation of order $\delta$: 
\begin{align*}
    & \log \frac{\norm{\vec{c}_{i,i+1} + \vec{\delta}_{i,i+1}}}{\norm{\vec{c}_{i,i+1}}} = \\
    &= \log \left(1 + \hat{c}_{i,i+1} \cdot \vec{\delta}_{i,i+1} + O(\delta^2) \right) \approx \hat{c}_{i,i+1} \cdot \vec{\delta}_{i,i+1}
\end{align*}
where $\hat{c}_{i,i+1} = \vec{c}_{i,i+1} / \norm{\vec{c}_{i,i+1}}$. Therefore, the ratio between the true scale factor and its noisy version is given by $e^{\tilde{\delta}}$, where $\tilde{\delta}$ is the random variable given by \Eq{log_scale_factor}. Its standard deviation scales as $\sqrt{j} \delta \leq \sqrt{n} \delta$, where $n$ is the total number of patches. While the order $\delta^2$ correction is always positive, resulting in a drift, it sums up across the patches to $O(n \delta^2)$, and is therefore higher order in $\sqrt{n} \delta$. Thus, as long as $\sqrt{n} \delta \ll 1$, the Lindbladian on each patch is stretched by a factor of at most $\approx 1 \pm \sqrt{n} \delta$, leading to a total recovery error of order $\sqrt{n} \delta$. This explains the square root scaling of the error with system size seen in \Fig{fig3}.



\end{document}